# An Optimized Combination of a Large Grid Connected PV System along with Battery Cells and a Diesel Generator

Farzad Ferdowsi[1], Ahmad Sadeghi Yazdankhah[2], Hossein Madadi Kojabadi[3]
[1]Electrical Engineering Department, Florida State University, Tallahassee, USA
[2,3]Electrical Engineering Department, Sahand University of Technology, Tabriz, Iran
[1]fferdowsi@fsu.edu

**Abstract**- Environmental, economical and technical benefits of photovoltaic (PV) systems make them to be used in many countries. The main characteristic of PV systems is the fluctuations of their output power. Hence, high penetration of PV systems into electric network could be detrimental to overall system performance. Furthermore, the fluctuations in the output power of PV systems make it difficult to predict their output, and to consider them in generation planning of the units. The main objective of this paper is to propose a hybrid method which can be used to control and reduce the power fluctuations generated from large grid- connected PV systems. The proposed method focuses on using a suitable storage battery along with curtailment of the generated power by operating the PV system below the maximum power point (MPP) and deployment of a diesel generator. These methods are analyzed to investigate the impacts of implementing them on the economical benefits that the PV system owner could gain. To maximize the revenues, an optimization problem is solved.

## Nomenclature

| | |
|---|---|
| B | Revenues gained from selling $P_G$ power over period T ($) |
| $C_D$ | Capital cost related to the generator power capacity ($/KW) |
| $C_E$ | Capital cost related to the battery energy capacity ($/KWh) |
| $C_M$ | Generator Fuel cost ($/Lit) |
| $C_P$ | Capital cost related to the battery power capacity ($/KW) |
| $C_{Pi}$ | Charges due to emission of harmful gases ($/Kg) |
| $E_b(i)$ | Battery energy in each section of the time (KWh) |

$E_{d_{Max}}$ Maximum energy produced by the diesel generator during the time study (KWh)

$i$ Number of time sections during the period of study

$m_i$ Amount of generated harmful gases due to diesel generator operation (Kg)

$N_B$ Number of times the battery will be replaced during the period T

$N_D$ Number of times the diesel generator will be replaced during period T

$O_D$ Operating and maintenance costs related to the generator power capacity ($/KW)

$O_E$ Operating and maintenance costs related to the battery energy capability ($/KWh)

$O_P$ Operating and maintenance costs related to the battery power capability ($/KW)

$P_b$ (i) Instantaneous battery power in each section of the time (KW)

$P_c$ (i) Curtailed power in each section of the time (KW)

$P_d$(i) Generator output power in each time step (KW)

$P_{bmax}$ Maximum power capability of the battery (KW)

$P_{dmax}$ Generator power capability (KW)

$P_G$ Injected power to the grid (KW)

$P_{G_{Max}}$ Maximum injected power to the grid (KW)

$P_{PV}$ (i) Instantaneous generated PV power (KW)

$R_L$ Consumed fuel for each KWh generated power (Lit)

$S$ Annual discount rate

$S_E$ Salvage value related to the battery energy ($/KWh)

$S_D$ Salvage value related to the generator power capability ($/KW)

$S_P$ Salvage value related to the battery power ($/KW)

$T$ Number of years studied

$T_B$ Life- time of the battery (years)

$T_D$ Life- time of the diesel generator (years)

$X_{min}$ Minimum admissible state of charge of battery

$\alpha$ Price of energy sold from the PV system ($/KWh)

$\beta$ Present worth corresponding to the battery power capacity ($/KW)

$\gamma$ Present worth corresponding to the battery energy capacity ($/KWh)

$\sigma$ Present worth corresponding to the diesel power capacity ($/KW)

$\rho_d$ Diesel generator efficiency

$\rho_P$ Power efficiency of the battery

$\rho_E$ Energy efficiency of the battery

## 1. Introduction

Photovoltaic Systems (PV) are used to convert solar energy to electricity through solar panels. Due to the environmental concerns, electricity productions based on green/clean fuel sources (such as sun) have become increasingly important. In recent years, grid- connected photovoltaic power plants of several megawatts have also seen commercial operations in many developed countries. In addition, PV systems have a number of technical and economical benefits, such as loss reduction and voltage profile improvement [1]. Unfortunately, instantaneous changes in the irradiance reaching the PV arrays and unpredictable nature of the solar power lead to a corresponding change in their output power. In most studies, the time frame for fluctuations in irradiance is in the order of few minutes especially suitable for systems with megawatts rating [2]. For large PV systems, the output power can change considerably in several minutes time frame, but for a number of small systems distributed over a large area, fluctuations are much less [3]. The frequency of these fluctuations are dictated by type and size of moving clouds [4], location and penetration level of the PV systems. As far as the penetration level is increased, the negative impacts of PV power fluctuations on some aspects of the network operation such as stability of the system, high variations in node voltages and scheduling of generation units would be increased [5-7].

The main target for utility is to have smooth output power of the PV system or to produce a certain limit of power from the PV systems. By applying certain technologies and methods, one could achieve to this task but, due to high costs of methods and the energy generated from the PV system, it is worthwhile to have economical analysis and then choose and implement suitable technology to smooth the power fluctuations.

In this matter, the use of energy storage systems (ESS) could be an effective method to mitigate these negative impacts [8] and [9]. In selecting a proper storage system one has to determine its applications. When a relatively high amount of power must be supplied in a very short period of time which is called power application; capacitors, super conducting magnetic storage systems and batteries with rapid reactions are vital [10]. But for energy application such as storing energy during the peak generation period and injecting this energy during the peak loading period, providing huge amount of energy during a certain period of time (hours up to days), hydro pump stations, hydrogen based storage systems and batteries usually are used. These mentioned technologies have different characteristics in performance. Hydro pump stations are not suitable for using with large PV systems, because they need large flat areas. Some of these technologies such as superconducting magnetic storage systems and ultra capacitors have very low energy densities and thus cannot be used alone in power applications [11, 12].

Categorizing batteries as the common member of both mentioned applications indicates that they can meet the required power capacity and necessary amount of energy. Moreover, batteries have quick response in both charge and discharge processes, so they are one of the most suitable technologies to reduce the output power fluctuations of PV systems.

When energy storage is considered, the storage reserve capacity and battery system capability has to be determined in terms of the days of battery back- up [13]. One approach for determination of battery capacity for a given system is to estimate it by multiplying the number of days that the system load could be supplied by the battery system with no power available from PV system. This method does not reflect the continuous variation of system generation and load on the storage capability. Also, it cannot be applied to small stand- alone power systems, because the stored energy is associated with only the generation shortage due to the solar resources unavailability. There are other methods that can be used to reduce power fluctuations. Installing a dump load will dissipate the excess PV power. The dump load consists of a resistance and a controller which controls the power flow through the load. In this method, dumping part of the generated power considerably will lead to loss of revenues. A third method is to curtail the power by operating below the maximum power point (MPP). This method only requires modification of the control strategy of the power conditioning unit but still the curtailed power shows considerable loss in revenues for all limits of power fluctuations.

This paper presents four different cases to implement the grid connected photovoltaic system using: 1) suitable battery storage, 2) battery storage system with curtailment of power by operating below the MPP, 3) battery storage and a back up diesel generator, and 4) combination of battery storage and a back up diesel generator with the curtailment of power by operating below the MPP. In all cases, the proper linear programming (LP) optimization problem with the corresponding constraints is applied. The problem is modeled in the LINGO software. The results obtained from solving the problem are compared in terms of costs and net benefit until the most efficient and economical option is obtained.

## 2. PV Power Fluctuations Reduction Using Batteries

### 2.1. Selecting the Most Optimum Battery

Fig. 1 shows a typical power waveform for 24 hours period of time based on real temperature and irradiance data obtained from a metrology station in southwestern United States with 10 minutes resolution [14]. As the figure shows, there are some times in which the generated power changes considerably and the amount of these changes may be over 50%. Since these amount of changes affects the power network adversely as mentioned, power fluctuations reduction should be considered vital.

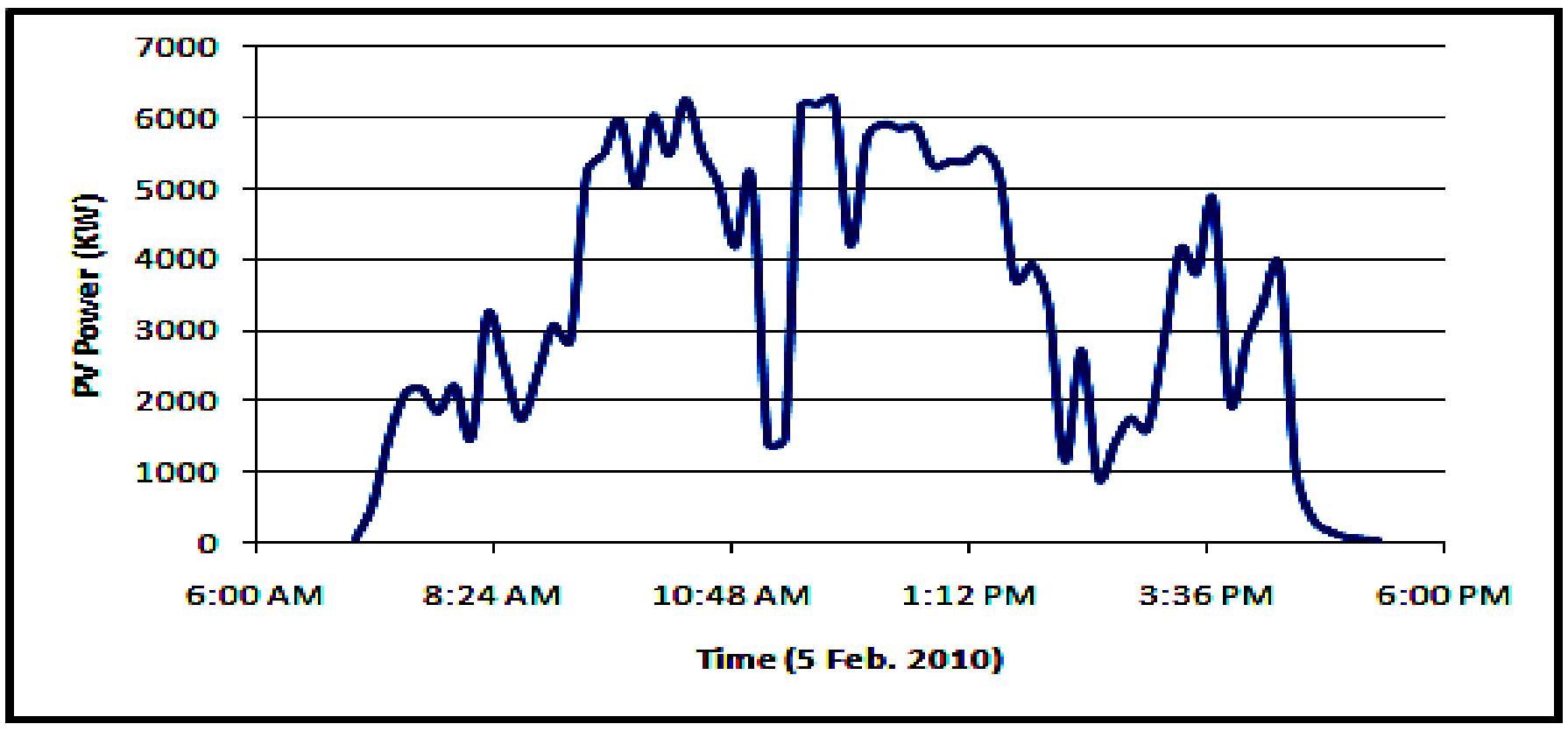

Fig. 1: PV power fluctuations

The appropriate amount of limitation on the power fluctuations of a grid connected DG system like PV is not constant for all circumstances and it is depended to several factors like required power quality, topology of the distributed network to which the DG system is connected, etc., but a reasonable value for this parameter is in the range of 0%-5% of the grid connected DG system capacity according to our simulations on several distributed network. The permissible fluctuation in the PV power fed to the grid is assumed to be $\pm 150$ kilowatts (1.5% of system capacity) per each 10 minutes time interval for all cases in this paper. Based on this demand, one has to use a suitable battery with the PV in order to satisfy the required operating conditions of the utility. In Ref. [15] mathematical analysis have been done in grid-connected photovoltaic/battery system shown in Fig. 2 to seek for the most optimum battery to control fluctuations.

.

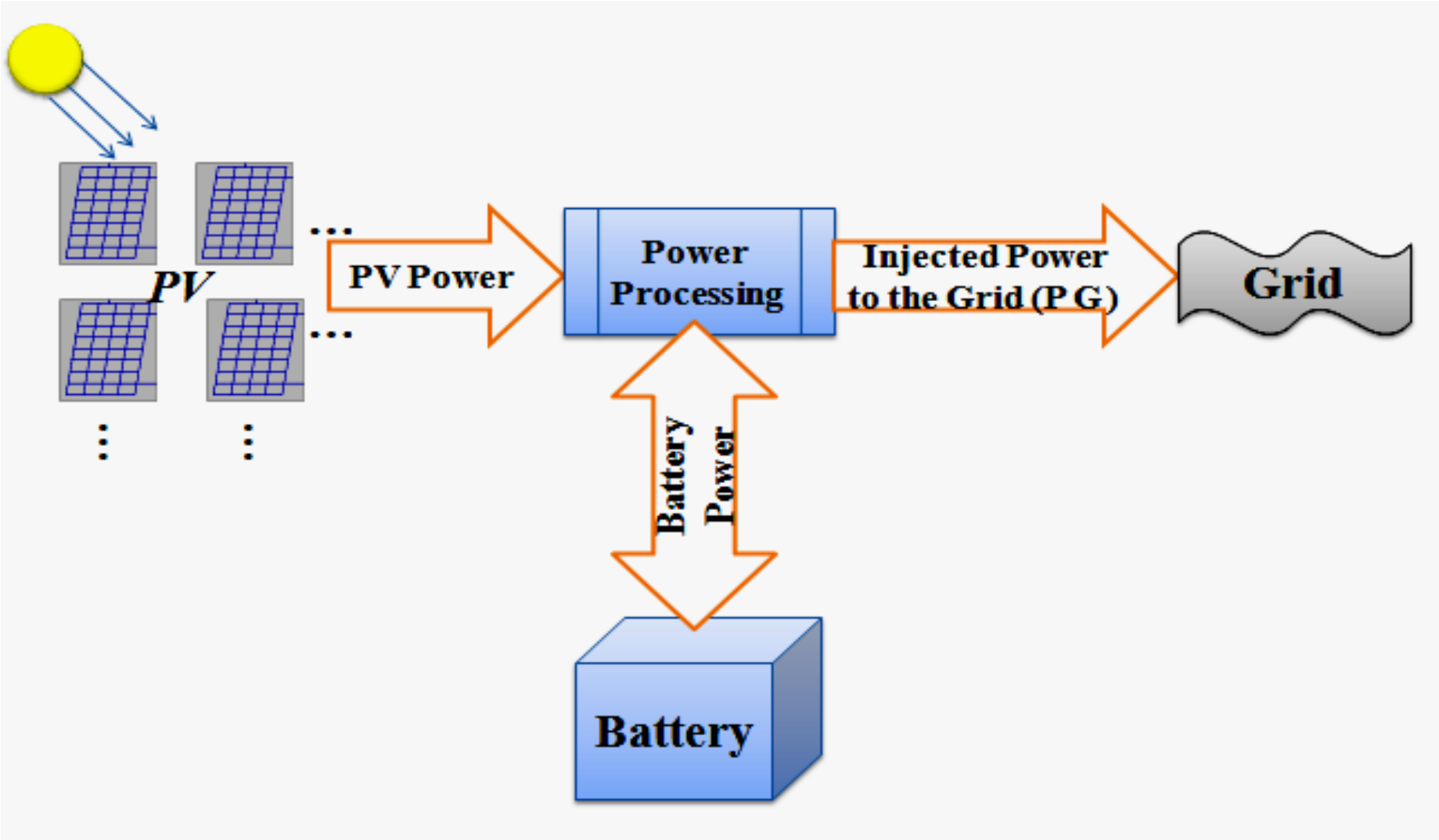

Fig. 2: Grid connected photovoltaic/battery system

Four ordinary batteries, Lead- Acid, Ni- Cd, NaS and Li- Ion, were studied technically in terms of their performance and then linear programming was applied as an optimization process to obtain the most economical option. The specifications of the four types of batteries are given in Table (1) [11], [16, 17]. Minimizing the cost function with the corresponding constraint used in Ref. [15] showed that NaS battery has the lowest cost for 18 years (period of study) assumption. Therefore, NaS battery is used for all cases in this paper as the most optimum battery .Table 2 represents the results of the battery optimization problem for the fluctuation constraint given in [15] .

Table 1: The specification of the four types of batteries

| | LA | Na-S | Li-ion | Ni-Cd |
|---|---|---|---|---|
| **Advantages** | Low capital cost | High power & energy ratings | High power & energy ratings, Low weight | Suitable power & energy ratings |
| **Disadvantages** | Low power & energy ratings | Production cost | Production cost, required special charging circuit | Harmful for environment |
| **$C_P$ $\frac{\$}{KW}$** | 300 | 1000 | 1300 | 600 |
| **$C_E$ $\frac{\$}{KWh}$** | 150 | 170 | 500 | 390 |
| $\rho_P$ | %85 | | | |
| $\rho_E$ | %85 | | | |
| $S_P$ $\frac{\$}{KW}$ | 3 | 10 | 13 | 6 |
| $S_E$ $\frac{\$}{KWh}$ | 1.5 | 1.7 | 5 | 3.9 |
| $O_P$ $\frac{\$}{KW}$ | 30 | 3 | 2 | 4 |
| $O_E$ $\frac{\$}{KWh}$ | 15 | 1.5 | 1 | 2 |
| $T_B$ years | 2 | 6 | 9 | 3 |
| $X_{MIN}$ | %10 | | | |

Table 2: Results of battery optimization problem

| | Net benefit ($\times 10^7$) $ | Imposition of further cost % (caused by using batteries) | Battery Power Rating (KW) | Battery Energy Rating (Kwh) |
|---|---|---|---|---|
| **LA** | 5.5 | -29.4% | 3963.031 | 9080.169 |
| **Na-S** | 6.2 | **-20.5%*** | 3604.084 | 9872.172 |
| **Ni-Cd** | 4.3 | -44.8% | 4192.814 | 8672.548 |
| **Li-ion** | 5.9 | -24.3% | 3600.104 | 9897.315 |

## 2.2. Control of Power Fluctuations Using Battery Storage System

In order to control the power fluctuations, a 10 MW grid connected PV with a BS system is considered as shown in Fig. 2. The permissible range of fluctuations is chosen $\pm$150KW/10min. To comply the power fluctuations limit, NaS battery is used which has better power and energy rating and minimum cost and maximum net benefit (Table 2). PV systems are allowed to inject variable generated power into the grid as long as the operating limits of the network are not violated. To control the output of the PV system, a BS system is included. To examine the economical aspect of installing BS system one could estimate the maximum revenues that the system owner can obtain. An optimization algorithm which is brought up in Ref. [11] is applied with slight changes to calculate this revenues as well as battery power and energy ratings. The input to the optimization process are time series of PV power with no battery included. The objective function (B) with used battery expenses $\beta$ and $\gamma$ are given as follows. It should be noticed that the selling price of energy is assumed to be 0.45 $/KWh and the time study is 18 years:

$$B = \sum_{k=1}^{T} \left[ (0.45 \times \frac{1}{6} \times \sum_{i} P_G(i)) \times \frac{1}{(1+s)^k} \right] - \beta \frac{1}{\rho_P} P_{bMAX} - \gamma \frac{1}{\rho_E} E_{bMAX} \quad (1)$$

$$\beta = \sum_{i=1}^{N_B} C_P \frac{1}{(1+S)^{(i-1)T_B}} + O_P \frac{(1+S)^{T_B} - 1}{S(1+S)^{T_B}} - \sum_{i=1}^{N_B} S_P \frac{1}{(1+S)^{i\,T_B}} \quad (2)$$

$$\gamma = \sum_{i=1}^{N_B} C_E \frac{1}{(1+S)^{(i-1)T_B}} + O_E \frac{(1+S)^{T_B} - 1}{S(1+S)^{T_B}} - \sum_{i=1}^{N_B} S_E \frac{1}{(1+S)^{i\,T_B}} \quad (3)$$

$$P_G(i) = P_{PV}(i) + P_b(i) \quad \forall i \quad (4)$$

$$|P_G(i) - P_G(i-1)| \leq 150 \quad \forall i \quad (5)$$

$$E_b(i) = E_b(i-1) - \frac{1}{6} P_b(i-1) \quad \forall i > 1 \quad (6)$$

$$E_b(i=1) = (r) \times E_{bMAX} \quad 0 \leq r \leq 1 \quad (7)$$

$$|P_b(i)| \leq P_{bMAX} \quad \forall i \quad (8)$$

$$(X_{MIN}\%) \times E_{bMAX} \leq E_b(i) \leq E_{bMAX} \quad \forall i \quad (9)$$

$$0 \leq P_G(i,j,k) \leq P_{GMAX} \quad \forall i, j, k \quad (10)$$

$$E_{bMAX}, P_{bMAX}, E_b, P_G \geq 0 \quad (11)$$

The first term in (1) calculates the revenues gained from selling energy to the grid. The second and third terms represent the net costs related to the battery power and energy ratings. In (2) and (3), the first term estimates the worth of capital cost related to battery power and energy ratings. The second term in both equations represents the worth of the annual operating and maintenance costs and the third term considers the salvage value at the end of battery lifetime. In (4) the injected power to the grid is calculated. Equation (5) ensures that power fluctuations are in the desired range. In (7), the initial state charge of battery is determined. Other constraints are given in (8) – (11).

## 3. Reduction of Power Fluctuations by Hybrid Control Method

The proposed method consists of a BS system as a main energy storage device along with a diesel generator and curtailment of the PV power by operating below the MPP as shown in Fig.3.

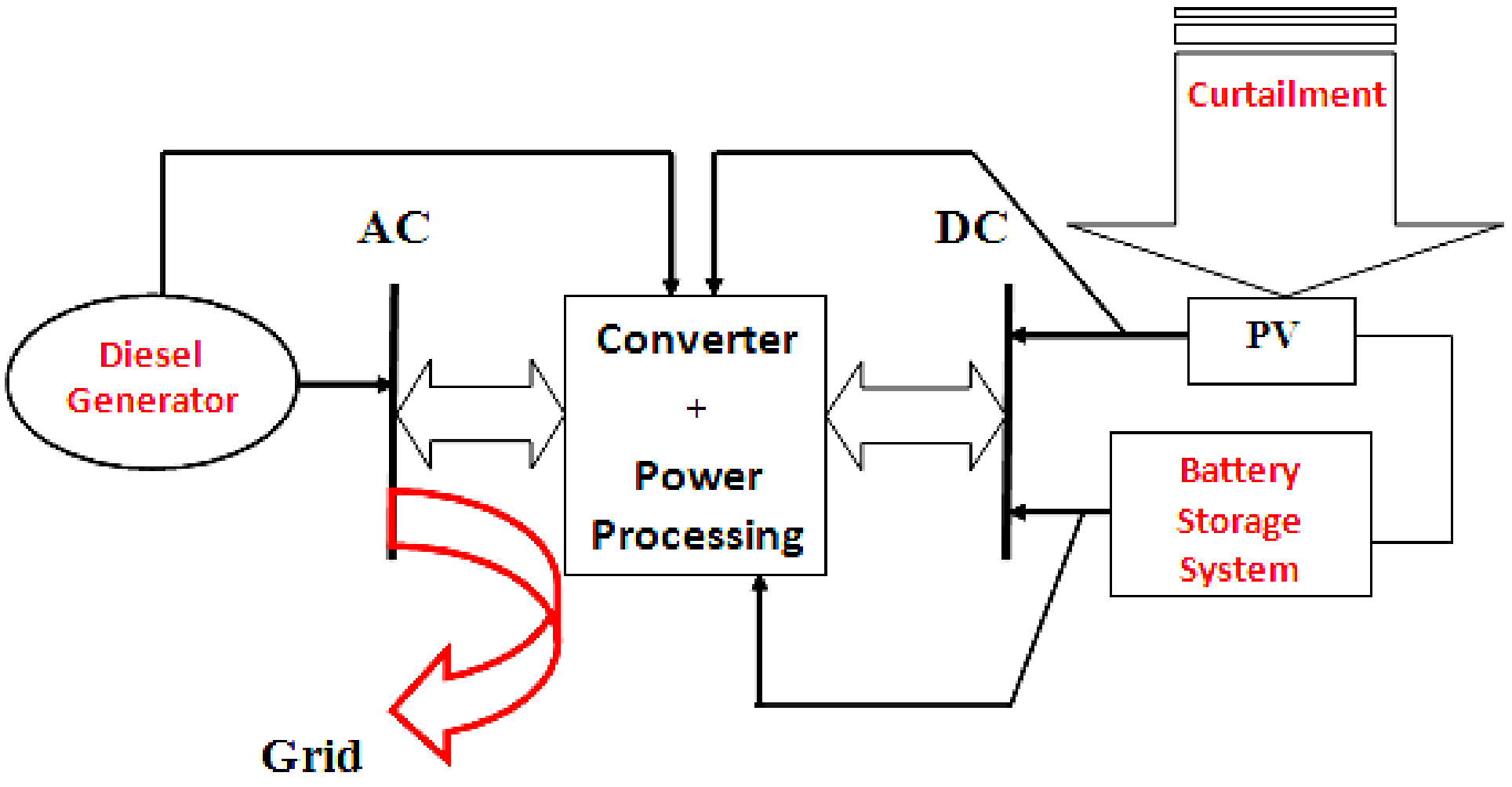

Fig. 3: Schematic of PV system with battery and diesel generator along with the curtailment based on operating below the MPP

Operating below the MPP control method acts as the BS system with only having charging process. The use of this method makes the system to operate in low level production. Integration of an external power supply such as diesel generator would solve this low level operation problem. Using the diesel generator along with the battery would reduce the size of battery (cost of battery) and increase its lifetime, because it responds very fast and has less expenses comparing to the battery.

The following relations with equations (12)-(16), represent the optimization problem for the proposed model.

$$B = \sum_{k=1}^{T}\left[(0.45 \times \frac{1}{6} \times \sum_i P_G(i)) \times \frac{1}{(1+s)^k}\right] - C_P \frac{1}{\rho_P} P_{bMAX} - C_E \frac{1}{\rho_E} E_{bMAX} - \sigma \frac{1}{\rho_D} \times P_{D_{Max}} - \left[\frac{1}{6} \times \sum_i P_D(i)\right] \times R_L \times C_M - [C_{P1} \times m_1 + C_{P2} \times m_2 + C_{P3} \times m_3] \quad (12)$$

$$\sigma = \sum_{i=1}^{N_D} C_D \frac{1}{(1+S)^{(i-1)T_D}} + O_D \frac{(1+S)^{T_D}-1}{S(1+S)^{T_D}} - \sum_{i=1}^{N_D} S_D \frac{1}{(1+S)^{iT_D}} \quad (13)$$

$$P_G(i) = P_{PV}(i) + P_b(i) - P_C(i) + P_D(i) \quad \forall i \quad (14)$$

$$0 \leq \left[\frac{1}{6} \times \sum_i P_D(i)\right] \leq E_{D_MAX} \quad (15)$$

$$E_{bMAX}, P_{bMAX}, E_b, P_G, P_D, E_{D_MAX} \geq 0 \quad (16)$$

(12) and (13) determines the system cost functions similar to equations (1) and (2) in the previous case. Equation (14) calculates the injected power to the grid in each section of the time. In (15) the operation of the diesel generator is limited by applying a specific constraint on its required fuel during the period of study. Finally, the last equation (16) is the non-negativity constraint.

4. Numerical Results

In this section, the linear programming based optimization problem is solved for the following cases and the results are shown and compared with each other. The case study is a 10 MW PV power plant that forms an area of about 310 square meters of land. The weather information such as temperature and irradiance is obtained with a resolution of ten minutes in an area with longitude and latitude 33.96 °N and 118.42 °W with an elevation of 32 meters above sea level in the United States [14]. The generated power from PV panels in each 10 minutes frame is estimated by means of the model described in [18]. Furthermore, the accuracy of this mathematical model has been verified for some weather data via the system that has been simulated in MATLAB/SIMULINK.

The study period is considered 18 years and it is assumed that changes in the climatic conditions for every year are almost the same. Thus, temperature and radiation are sampled during a year for each 10 minutes. Regardless of periods in which the radiation is less than 2 $W/m^2$, the number of annual output power samples is about 26000. Since the AC power should be injected to the grid, the inverter efficiency should also be taken into account. It is assumed the inverter's efficiency is constant and equal to 0.9. It should be also noticed that the permissible power fluctuation range is considered $\pm$150 KW/10 min for all cases.

### 4.1. Case A: PV Power Fluctuations Reduction Using BS System

As discussed earlier, the NaS battery was selected as the most optimum battery among four types of batteries. Therefore, NaS battery parameters are considered in solving the optimization problem. Results that come from solving the LP problem show that it is required to use a NaS battery with 2992.037 KW of power rating and 6012.457 KWh of energy rating. The results also indicate that using a NaS battery as the only method for fluctuation reduction reduce about 14.88% of the revenues that the owner of the PV system obtain in comparison with the case without any reduction in output power fluctuation. Figs. 4 and 5 show the original PV power before and after the fluctuation reduction as well as the battery power and energy rating for 3 consecutive days of January 2010.

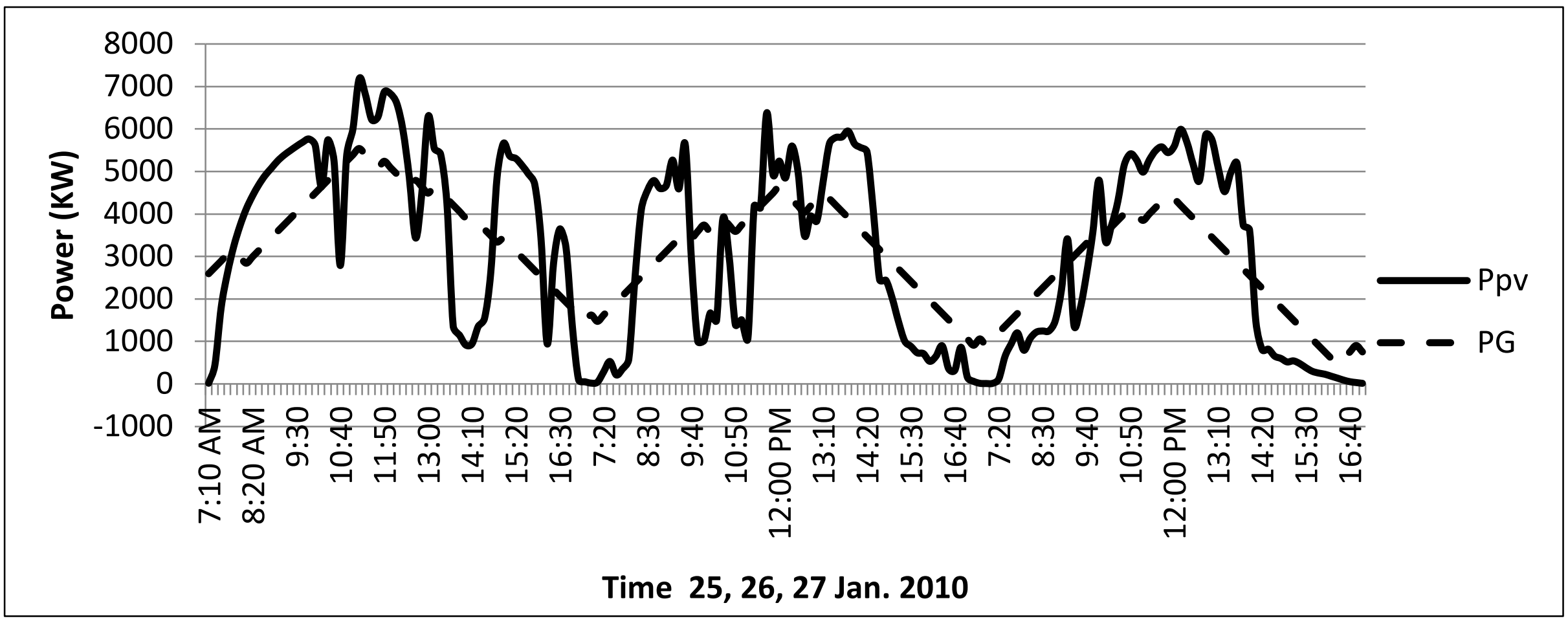

Fig. 4: Output power before and after the reduction of fluctuations

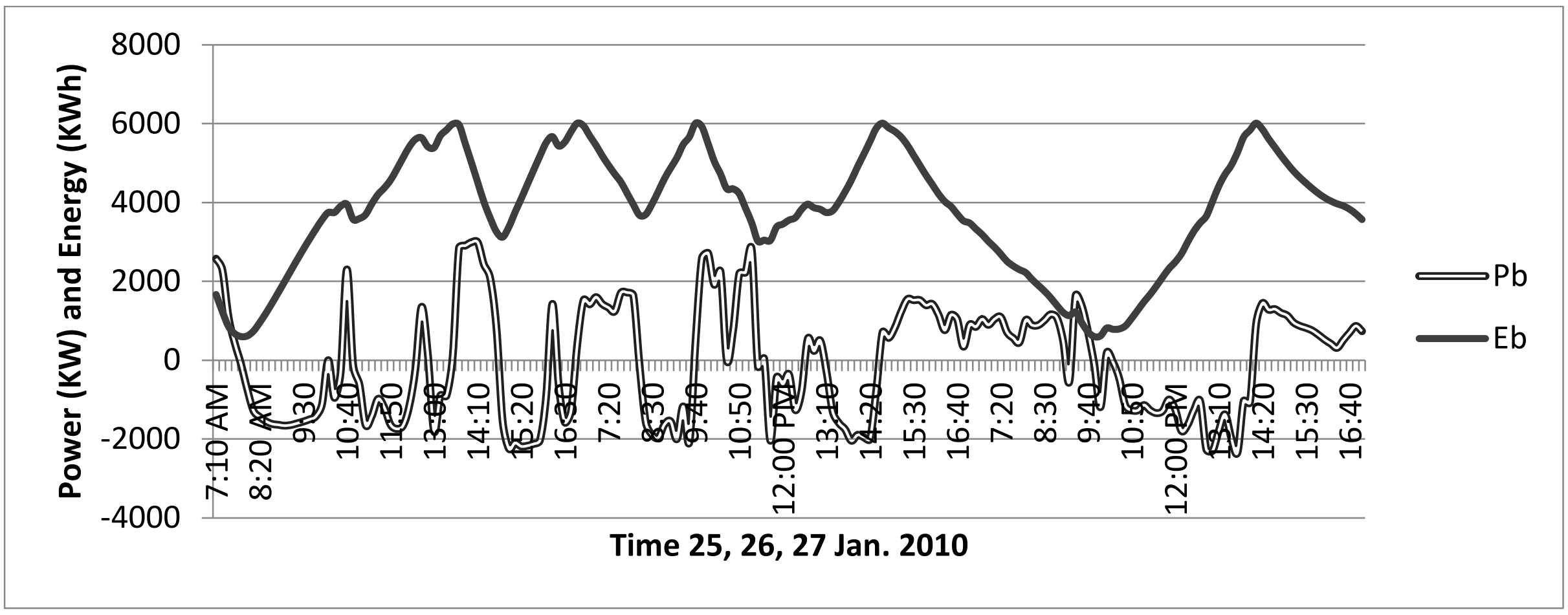

Fig. 5: Battery power (KW) and battery energy (KWh)

### 4.2. Case B: PV Power Fluctuation Reduction Using BS System Along with Operating PV below the MPP

In this case, NaS battery has been used as the base method for reducing the fluctuation. Obviously, in some periods there is excess power that should be injected to the battery to maintain the output power in the admissible range of fluctuation. Charging the battery specially in a short period of time not only reduce the life time of the battery, but also it can lead to demanding greater power and energy rating that would increase the size and cost of the battery. This concept could be led to an effective method by which we can prevent the PV system to generate that excess power. It may be done by operating the PV system below the MPP. Similarly, the required battery power and energy ratings as well as the curtailed power have been derived from solving the optimization problem (Solving equations (12) – (16) without considering the parameters related to the diesel generator). In this case, we need a NaS battery with 2354.679 KW of power rating and 4507.632 KWh of energy rating. The maximum curtailed power is also equal to 1497.947 KW.

Figs. 6 and 7 show the PV power before and after the reduction of fluctuations, battery power and energy and also the curtailed power.

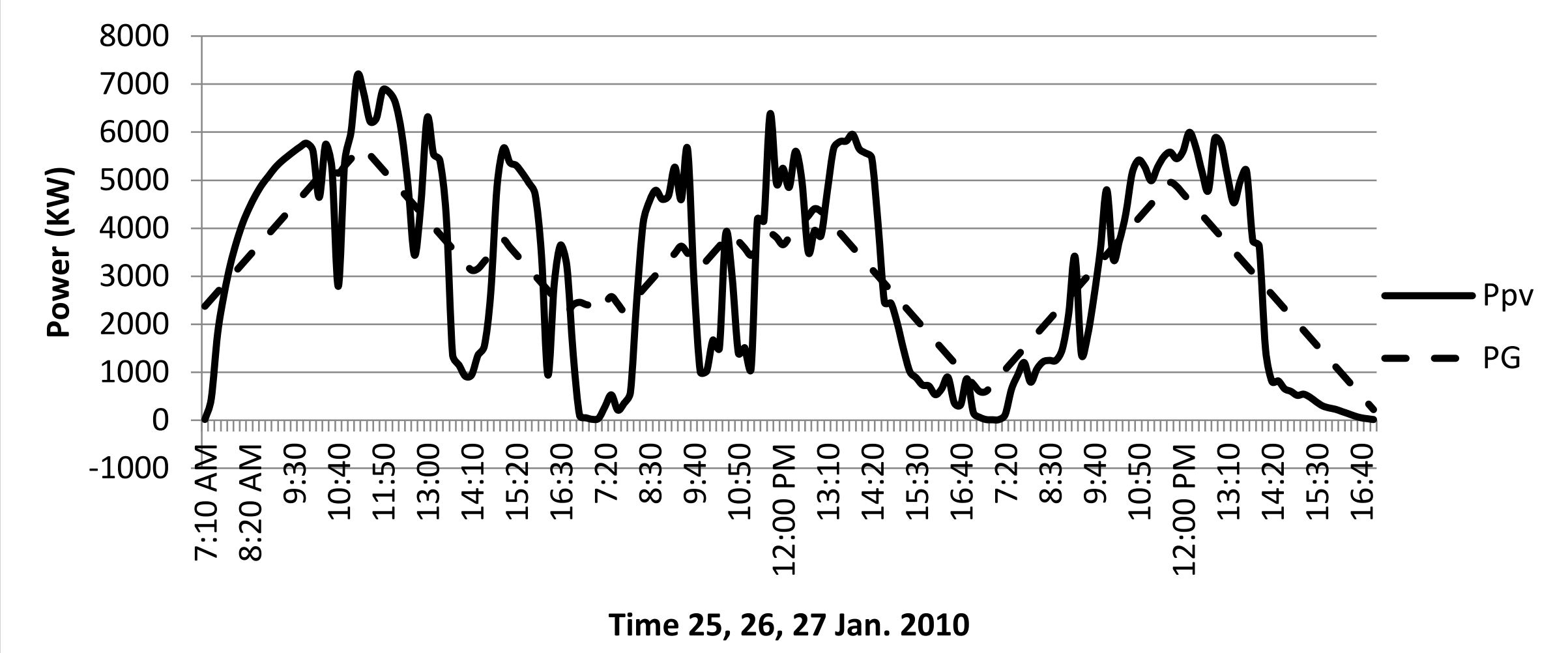

Fig. 6: Output power before and after the fluctuation reduction

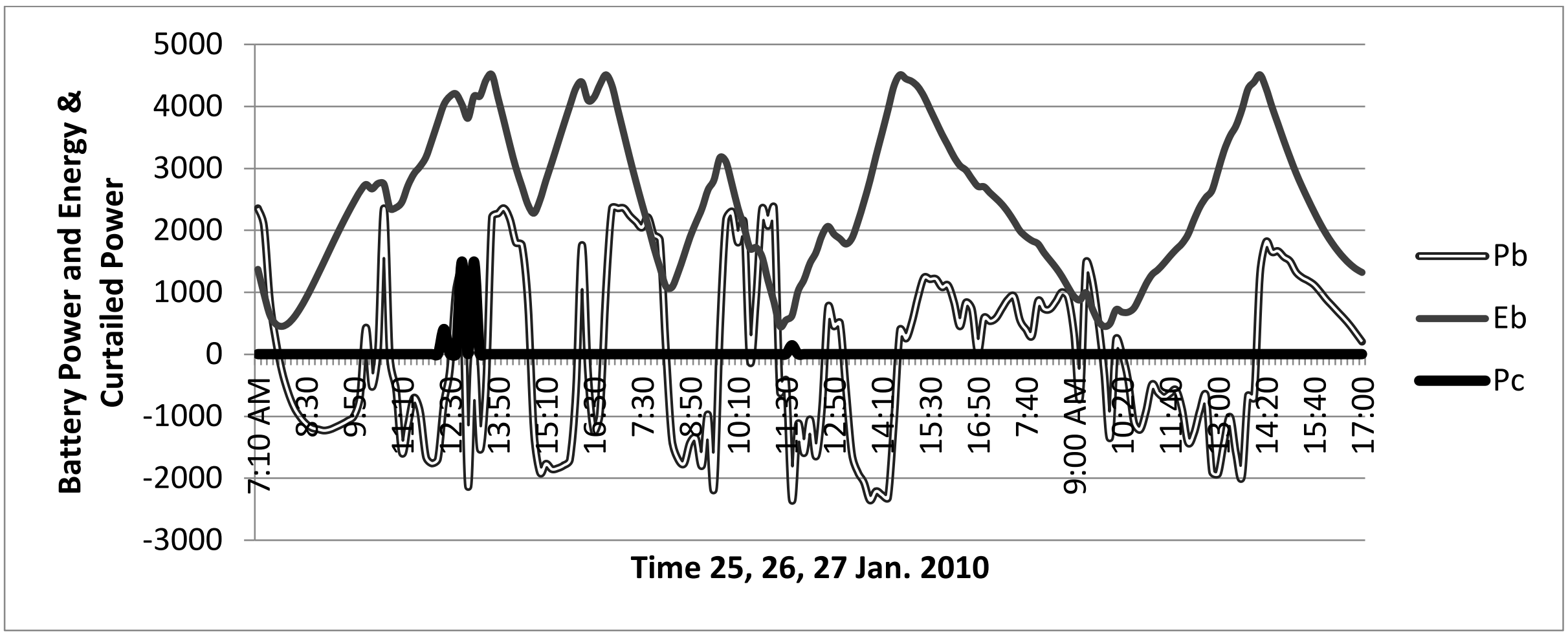

Fig. 7: Battery power, battery energy and curtailed power

### 4.3. Case C: PV Power Fluctuation Reduction Using BS System Along with a Diesel Generator

Since there are some periods of time that we may encounter shortage of power instead of having extra power, this case is somehow similar to the previous case. We have compensated the shortage of power in some periods by a diesel generator instead of the battery to increase the battery life time and reduce the battery size and cost. Characteristics of the employed diesel generator are given in Table (3) [19]. According to the lifetime of the generator, it should be replaced 3 times after initial installation during the period of study. Moreover, in the optimization problem, the fuel consumption of the diesel generator has been limited to 50000 liter per year (equation 15).

Table 3

| | | | |
|---|---|---|---|
| $C_D$ ($\$/_{KW}$) | 280 | $R_L$ ($^{Lit}/_{KWh}$) | 0.5 |
| $O_D$ ($\$/_{KW}$) | 80 | $C_M$ ($\$/_{Lit}$) | 1.1 |
| $S_D$ ($\$/_{KW}$) | 28 | $\sum C_{P_i} m_i$ (\$) | 6000 |
| $T_D$ **(hr)** | 20000 | **T (years)** | 18 |

In this case, the battery power rating is reduced to 1864.083 KW and its energy rating is also reduced to 3013.206 KWh. We will need a diesel generator with 1662.02 KW of power capacity. Figs. 8 and 9 show the output power, battery power and energy ratings and also diesel power profiles.

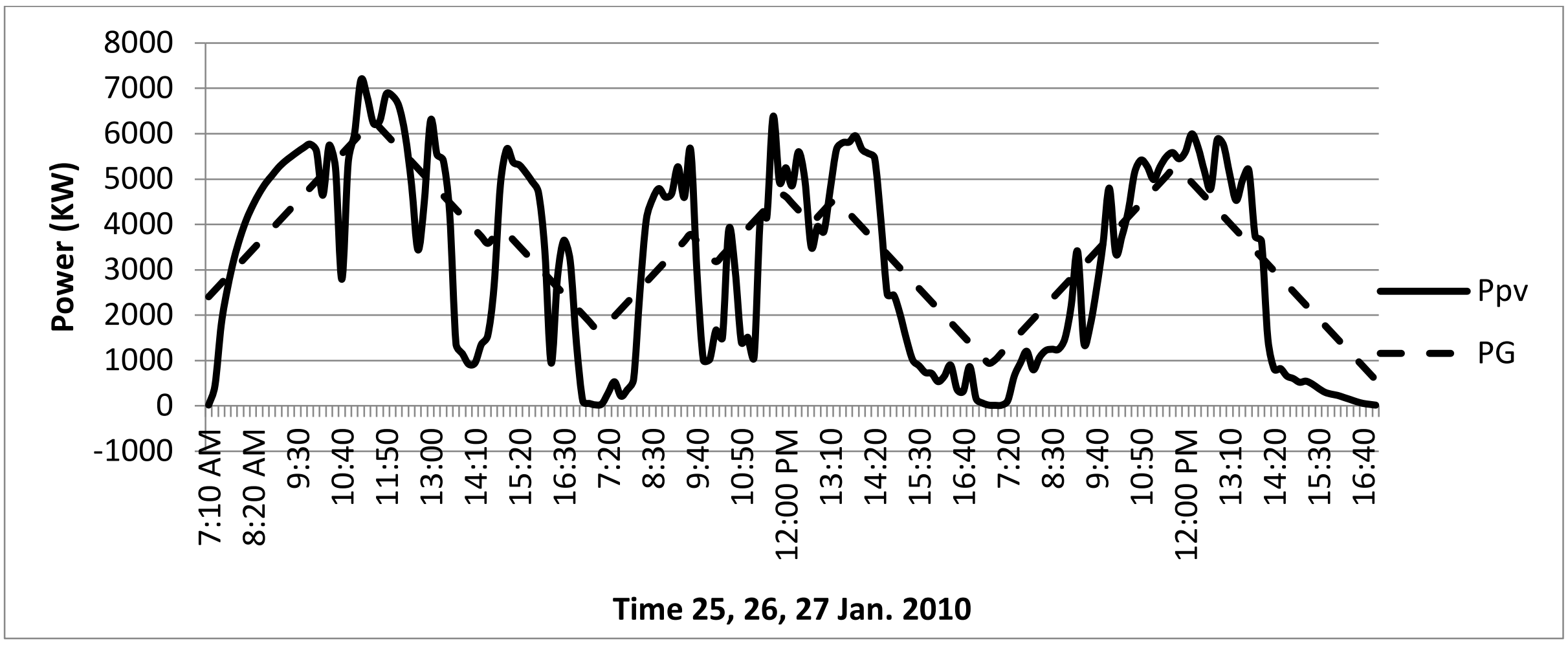

Fig. 8: Output power before and after applying the controlling method

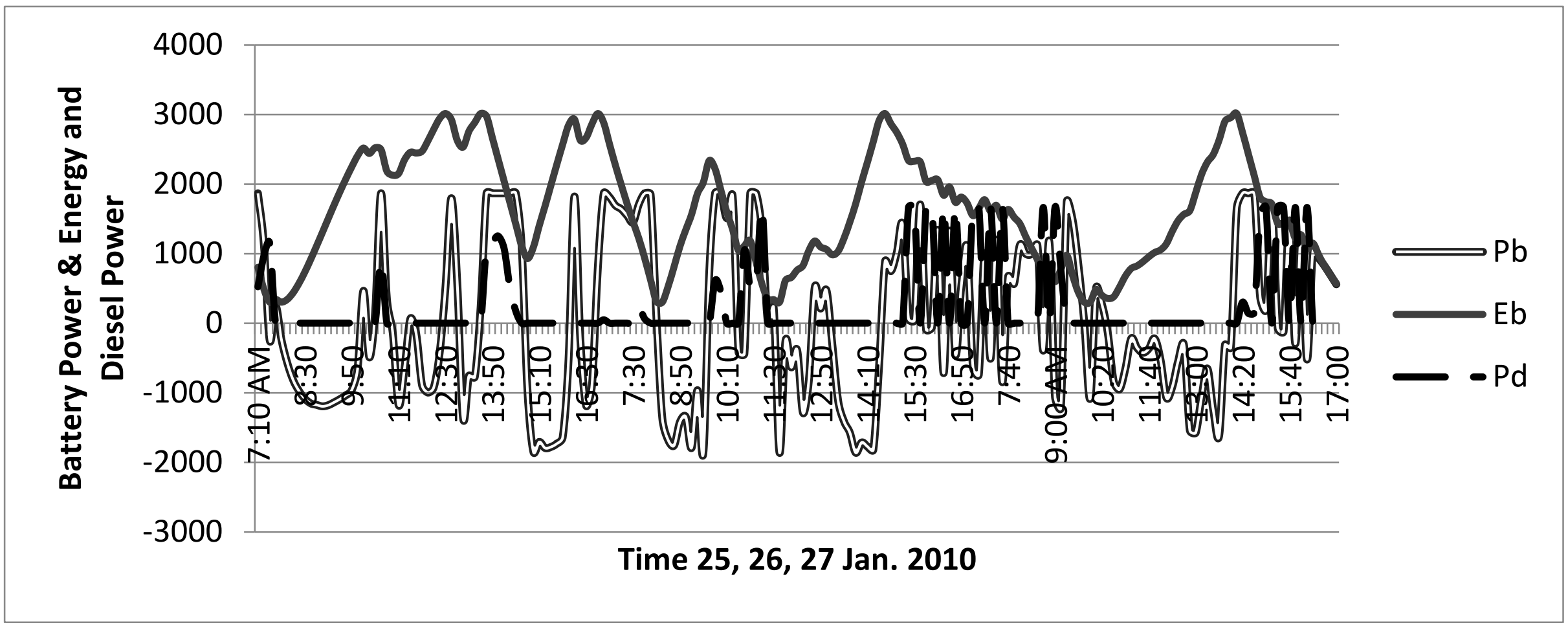

Fig. 9: Battery power, Battery energy and diesel generator power profiles

## 4.4. Case D: PV Power Fluctuation Reduction Using Hybrid Method (BS, Diesel Generator and Operating below the MPP)

In the last case, methods in case B and C have been combined together to reduce the cost and increase the revenues as often as possible. As a result, the size of the battery has been greatly decreased in comparison with the case A. Here a battery with 1335.165 KW of power rating and 2859.388 KWh of energy rating would be required that shows a great reduction in size and cost of the battery. In addition, the maximum curtailed power is equal to 2702.266 KW and the diesel power capacity is also 1532.335 KW. Figs. 10 and 11 show the diagrams of the output power, battery power and energy, curtailed power and diesel power profiles.

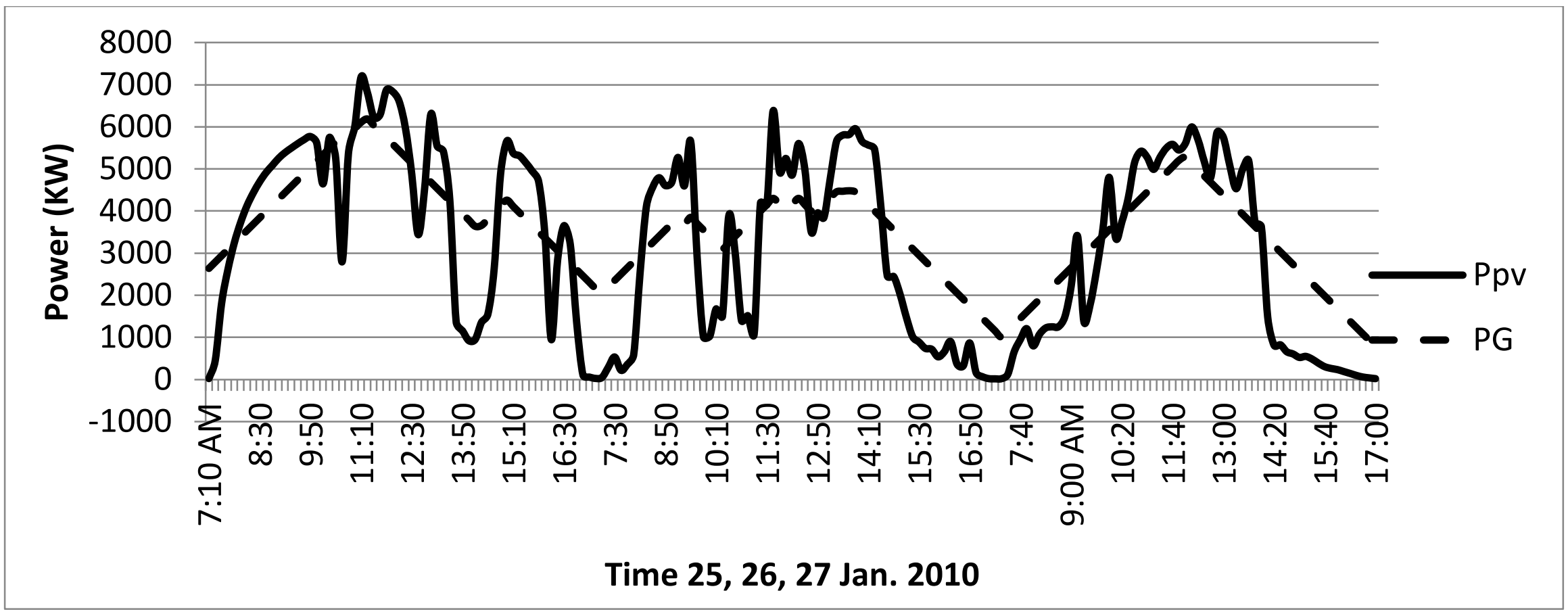

Fig. 10: Output power before and after of reduction of fluctuation

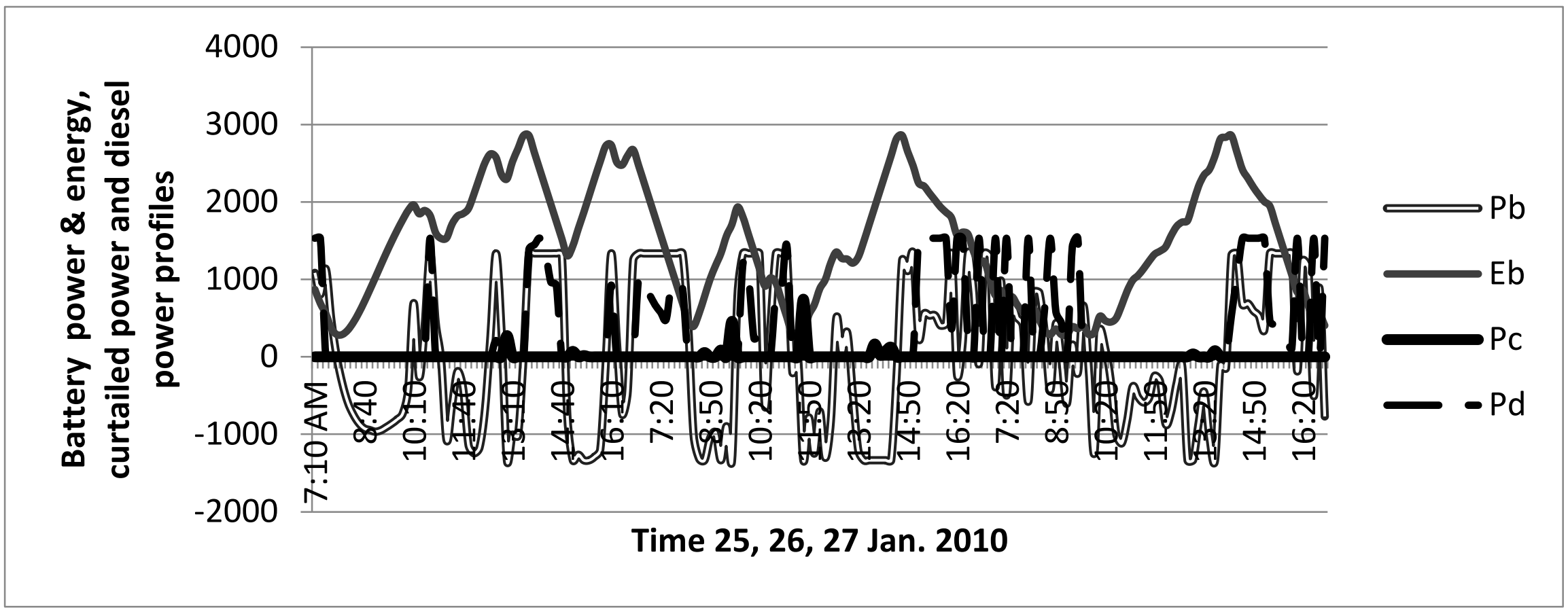

Fig. 11: Battery power & energy, curtailed power and diesel power profiles

Numerical results corresponding to each method are summarized in Table (4) for making a better evaluation of the four proposed methods.

Table 4: Numerical results of solving the LP optimization problem for the four proposed methods

| | Using NaS Battery | Battery + Curtailment | Battery + Diesel | Battery + Diesel + Curtailment |
|---|---|---|---|---|
| **Net revenues (×10$^{7}$) USD** | 6.39 | 6.81 | 6.88 | 6.99 |
| **Decrement in the revenues in comparison to the case with no fluctuation reduction tools** | -14.88% | -12.82% | -11.75% | **-10.38%*** |
| **Maximum curtailed power (KW)** | – | 1497.947 | – | 2702.266 |
| **Battery Power Rating (KW)** | 2992.037 | 2354.679 | 1864.083 | 1335.165 |
| **Battery Energy Rating (KWh)** | 6012.457 | 4507.632 | 3013.206 | 2859.388 |
| **Diesel Power Capacity (KW)** | – | – | 1662.020 | 1532.335 |

## 4. Conclusion

High penetration of PV system into electric network could be detrimental to overall system performance, because of having fluctuations in their output power. In this study, a hybrid method consisting suitable storage battery (NAS), backup diesel generator along with the curtailment of the generated power by operating below MPP was proposed to smooth the output power of the grid-interactive PV system. We focused on applying the most economical method to this system in order to decline the output power fluctuation velocity of the PVs connected to the grid. Four different cases were considered and in each case, accurate calculation and analysis of different aspects were included to have minimum cost and maximum benefits in terms of reducing battery size and expenses and the curtailed power of operation under MPP. Comparison of the results for different cases showed that in the area of study applying the combination of battery, power curtailment and diesel generator is the best method in terms of both technically and economically. It is clear that such studies can be performed in other areas, with regard to their whether circumstances.